\documentclass[11pt]{article}

\usepackage{amsthm,amsmath,amssymb,natbib}
\usepackage{graphicx,bm}
\usepackage{color}
\usepackage{lscape}
\usepackage{rotating}
\usepackage{setspace}
\usepackage[T1]{fontenc}
\usepackage{tgpagella}
\usepackage{dsfont}
\usepackage{bbold}
\usepackage{arydshln}
\usepackage[lined,boxed,commentsnumbered]{algorithm2e}
\usepackage[right=5cm]{geometry}                                        

\begin{document}

\begin{titlepage}
\title{The MELODIC family for simultaneous binary logistic regression in a reduced space}
\author{Mark de Rooij \\ Methodology \& Statistics Unit, Institute of Psychology, Leiden University\\ 
Patrick J. F. Groenen\\ Econometric Institute, Erasmus School of Economics, Erasmus University}
\date{\today}
\maketitle


\newpage

\begin{abstract}
Logistic regression is a commonly used method for binary classification.
Researchers often have more than a single binary response variable
and simultaneous analysis is beneficial because it provides insight
into the dependencies among response variables as well as between the
predictor variables and the responses. Moreover, in such a simultaneous analysis the equations can lend each other strength, which might increase predictive accuracy. 
In this paper, we propose the
MELODIC family for simultaneous binary logistic regression modeling. In
this family, the regression models are defined in a Euclidean space of
reduced dimension, based on a distance rule. The model may be
interpreted in terms of logistic regression coefficients or in terms of
a biplot. We discuss a fast iterative majorization (or MM) algorithm for parameter estimation. Two
applications are shown in detail: one relating personality
characteristics to drug consumption profiles and one relating
personality characteristics to depressive and anxiety disorders. 
We present a thorough comparison of our MELODIC family with alternative approaches for multivariate binary data.
\end{abstract}
\end{titlepage}

\section{Introduction}

Logistic regression \citep{berkson1944application, cox1958regression, agresti2003categorical} is one of
the most commonly used tools for binary classification. Although the
logistic function has been known since the early 19th century, the
logistic regression model was developed in the second half of the 20th
century \citep{cramer2002origins}. Adaptions of logistic regression models have
been developed to make it more flexible, through basis expansion,
or less flexible, by means of regularization. For an overview, see
\cite{friedman2001elements}.

Researchers regularly have more than a single response variable. That is, there are applications where several response variables can be predicted from a common set of predictor variables \citep{breiman1997predicting}. Examples include:

\begin{itemize}
\item The analysis of depressive and anxiety disorders. Mental disorders are highly prevalent in modern western societies and a high degree of comorbidity can often be observed among these disorders. In the Netherlands Study for Depression and Anxiety \citep{penninx2008netherlands} data were collected on a large number of subjects about their personality and about mental disorders \citep{spinhoven2009role}. 
\item The analysis of drug consumption profiles. \cite{fehrman2017five} are interested in subjects' drug consumption profiles and how these relate to personality characteristics such as sensation seeking and impulsivity. They collected data about the consumption of 18 different drugs. 
\item In clinical trials, the effect of treatments is established where the outcome can be dichotomous, cured or not cured. Treatments, however, come with side effects and these can be coded as present or absent. It is important to study treatment and side effects together in order to obtain the whole picture. For an empirical example, see \cite{molenberghs2006models}.
\item Psychosocial problems frequently occur in young adults. To screen for these problems in community settings, for example during large-scale general health check-ups, the Strengths and Difficulties Questionnaire (SDQ) can be used as it is a relatively short instrument. The SDQ has two parts: a self-report and a parent-report. To be useful as a screening tool it should have good validity properties, that is, it should be able to predict certain psychosocial problems. \cite{vugteveen2018using} investigated the validity of the SDQ with respect to four diagnoses. 
\end{itemize}

With multiple binary outcomes it is possible to fit a logistic regression model 
separately for each outcome, but it is often wise to build a single multivariate model.
In such a multivariate model the dependencies between the various outcomes can be better
understood and strength can be borrowed between the different outcomes. Second, such a multivariate model is more parsimonious in the sense that less parameters have to be estimated. 
Furthermore, estimated regression weights may be better in terms of mean squared
error. \cite{stein1956inadmissibility}, for example, showed that simple averages of a
multivariate normal distribution are inferior to shrunken averages in
terms of mean squared error, where the shrinkage tends toward the average
of averages. \cite{breiman1997predicting} show that shrinkage of coefficients of several multiple regression models toward each other is beneficial in terms of predictive accuracy. 
In a similar vein, building several logistic regressions in
a reduced space might provide better estimates of the regression
coefficients in terms of mean squared error.


There are basically two broad ways of analyzing multivariate data and
performing dimension reduction: The first is based on inner products from
which principal component analysis \citep{pearson1901principal, hotelling1936simplified, jolliffe2002principal} and reduced rank regression \citep{izenman1975reduced, braak1994biplots} are derived; the second is based on distances which have led to multidimensional
scaling \citep{torgerson1952multidimensional, torgerson1958theory, gower1966some, guttman1968general}
and multidimensional unfolding \citep{coombs1950psychological, roskam1968metric, heiser1981unfolding, busing2010advances}. This distance
framework is conceptually easier than the inner product framework  and leads to more straightforward interpretation \citep{derooij2005graphical}. Distances, especially Euclidean and Manhattan, are all around us and can already be understood by very young children.

In multidimensional unfolding, we generally have a dissimilarity matrix between two sets of objects. The goal is to find a low-dimensional mapping including points for the row objects and the column objects such that the distances between the points of the two sets are as close as possible (often in the ``least squares'' sense) to the observed dissimilarities. We will develop a family of models based on similar ideas. 

In this paper, we will develop a family of logistic models within a
distance framework. We call it the MELODIC family, written out, the
\underline{M}ultivariat\underline{E} \underline{LO}gistic
\underline{DI}stance to \underline{C}ategories family. 
More specifically,
we will develop a framework of models in which both participants and the
categories of the different response variables have a position in low-dimensional Euclidean space. The distances between the position of a
participant and the positions of the two categories of a single response
variable determine the probabilities for these two response options. The position of a participant will be parameterized as a linear combination of the predictor variables. 

The family extends the recently proposed multivariate logistic distance
models \citep{worku2018multivariate} which built on earlier logistic
distance models \citep{takane1987ideal, takane1987analysis, derooij2009ideal}
and can be considered as examples in the ``Gifi goes logistic''
framework as laid out by \cite{deleeuw2005gifi} and \cite{evans2014logistic}.

In the next section, we will develop the general model and two
constrained variants. We will discuss properties of the models and
provide interpretational rules. Two types of these rules can be
distinguished: the numerical and the graphical. These two modes of
interpretation for a single model are beneficial because there are those people who say that
``a graph is worth a thousand words'', the so-called  \emph{graph people} \citep{friendly2011comment}, while others (the \emph{table people}) firmly disagree \citep{gelman2011tables}.
In Section 3, we develop an Iterative Majorization or MM
algorithm \citep{deleeuw1977convergence, groenen1993majorization, heiser1995convergent, hunter2004tutorial}
for estimating the parameters of our models by minimizing a deviance function. Section
4 describes two illustrative applications. In Section 5, we discuss related statistical models and provide some comparisons. 
We conclude, in Section 6, with a
general discussion of our developments and some possibilities for further investigation.

\section{MELODIC family}\label{melodic-family}

\subsection{Data and notation}\label{data-and-notation}

We consider a system with \(P\) explanatory, predictor, or independent
variables \(X\) and \(R\) outcome, response, or dependent binary
variables \(Y\). That is, we have a sample of observations
\(\{\mathbf{x}_i,\mathbf{y}_i\}_{1}^n\) with
\(\mathbf{x}_i \in \mathbb{R}^P\) and \(\mathbf{y}_i \in \{0,1\}^R\).
The response variables will be recoded into indicator vectors \(\mathbf{g}_{ir}\) of length two, where the first element equals 1 if  \(y_{ir} = 0\) and the second element equals 1 if  \(y_{ir} = 1\).

We will use the following notation.

\begin{itemize}
\item
  \(i = 1,\ldots,n\) for individuals (participants, subjects, objects).
\item
  \(p = 1,\ldots,P\) for predictor variables (explanatory or independent variables).
\item
  \(r = 1,\ldots,R\) for response variables (outcome or dependent variables).
\item
  \(m = 1,\ldots,M\) an indicator for the dimensions.
\item
  There is a set of predictor variables \(X = \{X_p\}_{p=1}^P\). Observed values of the predictor variables are
  collected in the \(n \times P\) matrix \(\mathbf{X}\) with elements
  \(x_{ip}\). We assume, without loss of generality, that the predictor
  variables are centered, that is
  \(\mathbf{1}^\top\mathbf{X} = \mathbf{0}\).\\
\item
  There is a set of response variables \(Y =\{Y_r\}_{r=1}^R\). Observed values of the response variables are
  collected in the \(n \times R\) matrix \(\mathbf{Y}\). The matrix has
  elements \(y_{ir} \in \{0,1\}\). We will code the responses in a 
  super indicator matrix \(\mathbf{G}\) having \(C = 2R\) categories,
  that is, \[
  \mathbf{G} = [\mathbf{G}_1| \mathbf{G}_2| \ldots | \mathbf{G}_R].
  \]
\item
  \(\mathbf{B}\) represents a \(P \times M\) matrix with regression weights for the
  predictor variables.
\item
  \(\mathbf{u}_i\) is an \(M\) vector with coordinates for person \(i\) in
  \(M\)-dimensional Euclidean space. These coordinates will be collected
  in the \(n \times M\) matrix \(\mathbf{U}\) with elements \(u_{im}\).
\item
  \(\mathbf{V}_{r}\) is a \(2 \times M\) matrix having the coordinates of category
  \(0\) (i.e., \(\mathbf{v}_{r0}\)) in the first row and in the second row the coordinates of category
  \(1\) (i.e., \(\mathbf{v}_{r1}\)), both for response variable \(r\). These matrices will be collected in the \(2R \times M\) matrix
  \(\mathbf{V} = \left[\mathbf{V}^\top_1, \ldots, \mathbf{V}^\top_R\right]^\top\)
  with elements \(v_{rcm}\).
\item
  The observations are \(\{\mathbf{x}_i, \mathbf{y}_i\}_1^n\).
\item
  We define a block diagonal matrix \(\mathbf{J}\) with \(2 \times 2\) diagonal blocks
  \(\mathbf{I}_2 - \frac{1}{2}\mathbf{11}^\top\).
\item
  We use tildes for current estimates in the iterative process, that is,
  \(\widetilde{\mathbf{B}}\) represents the matrix with estimates in a
  given cycle of the algorithm.
\item
  Diag() denotes the operator that takes the diagonal values of a matrix
  and places them in a vector.
\end{itemize}

\subsection{General model}\label{general-model}

We define the conditional probability that person \(i\) is in class \(c\)
(\(c = \{0,1\}\)) of response variable \(r\),
\(\pi_{rc}(\mathbf{x}_i) = P(Y_{ir} = c|\mathbf{x}_i)\) as 
\begin{eqnarray}
\pi_{rc}(\mathbf{x}_i)  = \frac{\exp(-\delta(\mathbf{u}_i, \mathbf{v}_{rc}))}{\exp(-\delta(\mathbf{u}_i,\mathbf{v}_{r0})) + \exp(-\delta(\mathbf{u}_i,\mathbf{v}_{r1}))},
\label{eq:model}
\end{eqnarray} 
where \(\delta(\cdot, \cdot)\) denotes half the squared Euclidean
distance 
\begin{eqnarray}
\delta(\mathbf{u}_i,\mathbf{v}_{rc}) = \frac{1}{2}\sum_{m=1}^M \left(u_{im} - v_{rcm} \right)^2 = \frac{1}{2}\sum_{m=1}^M \left(u_{im}^2 + v_{rcm}^2 - 2u_{im}v_{rcm} \right),
\label{eq:dist}
\end{eqnarray}
 in \(M\)-dimensional Euclidean space. The dimensionality \(M\) has to
be chosen by the researcher with possible values being between 1 and
\(\min(P,R)\). The coordinates of the subjects (\(\mathbf{u}_i\)) are
assumed to be a linear combination of the predictor variables, that is,
\(\mathbf{u}_i = \mathbf{x}_i^\top\mathbf{B}\), where \(\mathbf{B}\) is
a \(P \times M\) matrix with regression weights. The coordinates of
category \(c\) of response variable \(r\) on dimension \(m\) are denoted
by \(v_{rcm}\) and collected in the \(M\)-vector \(\mathbf{v}_{rc}\).

Every subject \(i\) is thus represented in an \(M\)-dimensional Euclidean
space. Moreover, this subject has a distance to a point representing category 1 of
response variable \(r\) and to a point representing category \(0\) of response variable
\(r\). These two distances determine the probability for the subject to
answer with either of these categories; the smaller the distance, the larger the probability. In other words, a subject is most
likely to be in the closest class.

The log odds in favor of the 1 category and against category 0 for
response variable \(r\) given the subject's position is given by 
\begin{eqnarray}
\log{\frac{\pi_{r1}(\mathbf{x}_i)}{1 - \pi_{r1}(\mathbf{x}_i)}} = \log{\frac{\pi_{r1}(\mathbf{x}_i)}{\pi_{r0}(\mathbf{x}_i)}} = \delta(\mathbf{u}_i,\mathbf{v}_{r0}) - \delta(\mathbf{u}_i,\mathbf{v}_{r1}),
\label{eq:logodds}
\end{eqnarray} 
a simple difference of squared Euclidean distances. This log
odds can be further worked out as 
\begin{eqnarray}
\log{\frac{\pi_{r1}(\mathbf{x}_i)}{\pi_{r0}(\mathbf{x}_i)}} =\sum_{m=1}^M  \left[ \frac{1}{2}(v^2_{r0m} - v^2_{r1m}) + \mathbf{x}^\top_i\mathbf{b}_m (v_{r1m} - v_{r0m}) \right],
\label{eq:logodds2}
\end{eqnarray}
where we see that the effect of predictor variable \(p\) on response
variable \(r\) is determined by the regression coefficients \(b_{pm}\)
and the distance between the two categories. In general, the further apart the
two categories are, the better they are discriminated by the
predictor variables. If the two categories fall on the same position in
the Euclidean space, they are \textit{indistinguishable} \citep{anderson1984regression} based on
this set of predictor variables.

Let us define
\(a^*_r = \frac{1}{2}\sum_{m=1}^M (v^2_{r0m} - v^2_{r1m})\) and
\(\mathbf{b}^*_r = \sum_{m=1}^M \mathbf{b}_m (v_{r1m} - v_{r0m})\). Then the log odds can be written as 
\begin{eqnarray}
\log{\frac{\pi_{r1}(\mathbf{x}_i)}{\pi_{r0}(\mathbf{x}_i)}} = a^*_r + \mathbf{x}^\top_i\mathbf{b}^*_r,
\label{eq:lrcoefs}
\end{eqnarray}
showing that the model can be interpreted as standard binary logistic regression models.
We call the $b^*$ the model \textit{implied coefficients}. 

Equation \ref{eq:lrcoefs} also shows that our model is equivalent to a logistic reduced rank regression model. The matrix of regression coefficients with columns $\mathbf{b}^*_r$ has a rank $M$ constraint and can be decomposed in a matrix with $\mathbf{b}_m$ and a matrix containing the differences $v_{r1m} - v_{r0m}$. The logistic reduced rank regression is a special case of reduced rank vector generalized linear models as has been proposed by \cite{yee2003reduced}. 

\subsection{Constrained models}\label{constrained-models}

In the general model described above, the categories of the response
variables lie freely somewhere in the \(M\)-dimensional space.
Sometimes, however, researchers already have an idea about the
underlying structure of the response variables. In the literature about
depressive and anxiety disorders, for example, one theory says
that fear and distress are its underlying dimensions, where each dimension comprises a subset of the
disorders. In terms of our models, this means that the categories of the
response variables pertaining to the distress dimension lie on a single
dimension (i.e., ~the coordinates of the categories of these response for the other dimensions equal zero).
Similarly, for the categories of the response variables
pertaining to the fear dimension, the coordinates on the distress
dimension all equal zero.

If a specific response variable pertains to, say, dimension 1, the class coordinates on all other dimensions are set to zero, that is $v_{r1m} = v_{r0m} = 0, \forall m \neq 1$. 
Such a structure simplifies the model and its interpretation, because in
the log odds definition (see Equation \ref{eq:logodds2}) the last term becomes zero for several dimensions and only the
regression weights of the dimension to which the response variable
pertains are important for the discrimination of the categories of that
response variable.

One further constraint is to let all response variables have the same
discriminatory ability. In that case, \((v_{r1m} - v_{r0m}) = 1\) for the
dimensions to which response variable \(r\) pertains. For this constrained model, Worku and De Rooij (2018) showed that the parameters can be estimated using standard software for logistic regression by using a structured design matrix for the predictors. This paper
describes them as members of a larger family of models, the MELODIC family.

\subsection{Graphical representation}\label{graphical-representation}

When the dimensionality equals two (\(M=2\)), the model can be easily represented
graphically. This representation shows 1) the categories of
the response variables as points, 2) a
decision line for every response variable designating the predicted
class at a specific point, 3) variable axes for the predictor variables, and 4) the subjects'
positions as points. Many aspects of the
interpretation of these graphical representations follow the theory of
biplots as discussed in \cite{gower1996biplots} and \cite{gower2011understanding}.

\begin{figure}
\begin{center}
\includegraphics[width = .9\textwidth]{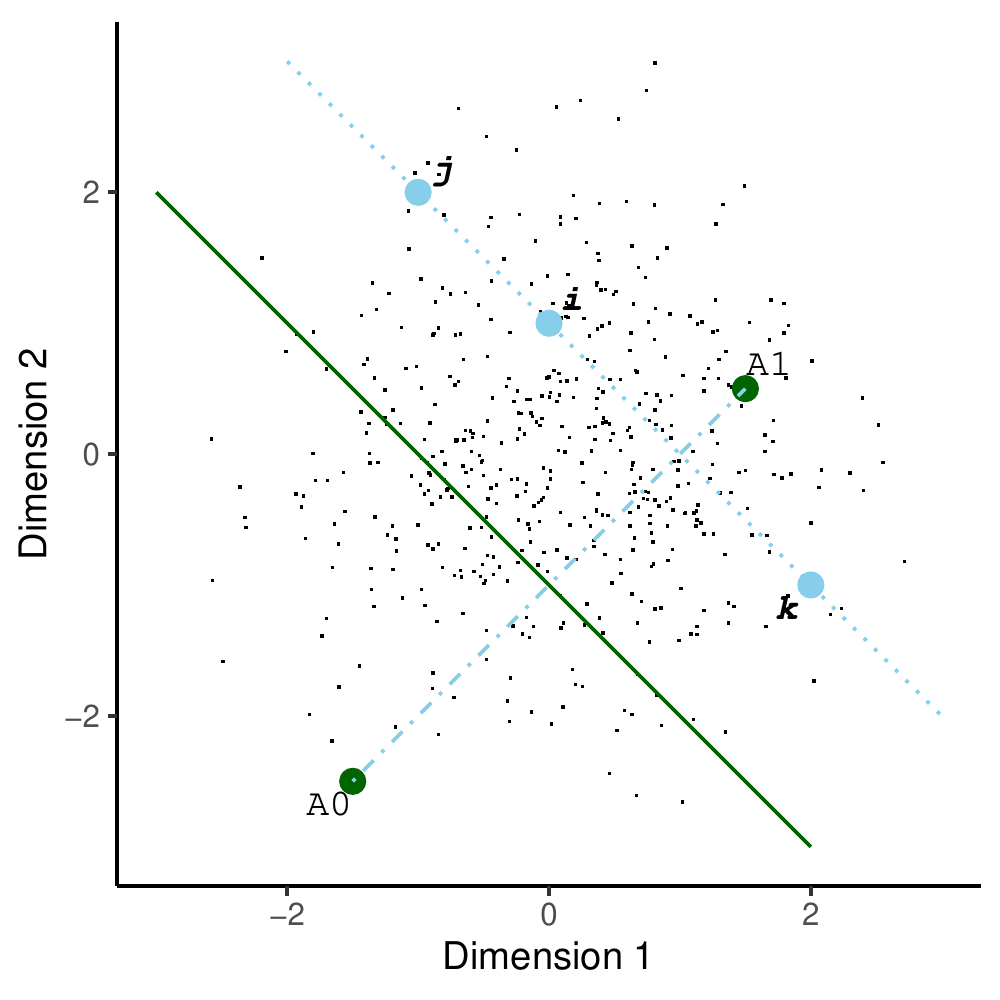}
\caption{Graphical representation with a single dichotomous response variable A (with categories A0 and A1) and three participants ($i$, $j$, and $k$). The solid green line represents the decision line where the probabilities for A0 and A1 are equal. The blue dotted line projects the points representing the three subjects onto the A01 line (blue dashed-dotted line). All points on this dotted line represent observations with the same log odds.}
\label{fig:fig1}
\end{center}
\end{figure}

Let us first look at a graphical representation for a single response
variable and a set of subjects, of which three of them are highlighted. Figure
\ref{fig:fig1} gives such a graph where A0 and A1 are the two categories of a
response variables named A, and $i$, $j$ and $k$ present three subjects. The line halfway between
classes A0 and A1 represents the decision line, in other words the line
represents the points for which the odds are even. The log odds that
subject $i$ chooses A0 instead of A1 are clearly in favor of class A1, because
that is the closest class. 

The squared distances from Subject $i$ to
categories A0 and A1 additively decompose into one part toward the line through A0 and A1 (i.e., the A01 line)
and one part along this line.  Equation \ref{eq:logodds}
shows that the log odds are defined in terms of a difference in squared distances, and therefore the
part toward the A01 line drops out of the equation. In more detail, for this example we have
\begin{eqnarray*}
\log{\frac{\pi_{A0}(\mathbf{x}_i)}{\pi_{A1}(\mathbf{x}_i)}} &=& \delta(\mathbf{u}_i,\mathbf{v}_{A1}) - \delta(\mathbf{u}_i,\mathbf{v}_{A0}). 
\end{eqnarray*}
According to the Pythagorean theorem, the squared distance $ \delta(\mathbf{u}_i,\mathbf{v}_{A1})$ can be decomposed into $ \delta(\mathbf{u}_i,\mathbf{v}_{A01}) +  \delta(\mathbf{v}_{A01},\mathbf{v}_{A1})$, where $\mathbf{v}_{A01}$ is the coordinate of the projection of $\mathbf{u}_i$ on the A01 line. Using this decomposition for both terms we obtain
\begin{eqnarray*}
\log{\frac{\pi_{A0}(\mathbf{x}_i)}{\pi_{A1}(\mathbf{x}_i)}} &=& 
\left(\delta(\mathbf{u}_i,\mathbf{v}_{A01}) +  \delta(\mathbf{v}_{A01},\mathbf{v}_{A1})\right)  - 
\left(\delta(\mathbf{u}_i,\mathbf{v}_{A01}) +  \delta(\mathbf{v}_{A01},\mathbf{v}_{A0})\right)  \\
&=&  \delta(\mathbf{v}_{A01},\mathbf{v}_{A1}) - \delta(\mathbf{v}_{A01},\mathbf{v}_{A0}).
\end{eqnarray*}
For person $j$ or $k$, we can use the same decomposition. The projections for the three subjects are however equal, and therefore the log
odds of category A0 against A1 for persons $i$, $j$, and $k$ are equal (and for all three subjects in favor of category A1). As noted above, the decision line represents the set of positions where the odds are even, that is, the log odds are equal to zero. More generally, iso log odds curves, which are curves where the log odds equal any constant, are straight lines parallel to these decision lines and orthogonal to the A01 line. An example of such an iso log odds line is the one through the points representing the three subjects (blue dotted). 

\begin{figure}
\begin{center}
\includegraphics[width = .9\textwidth]{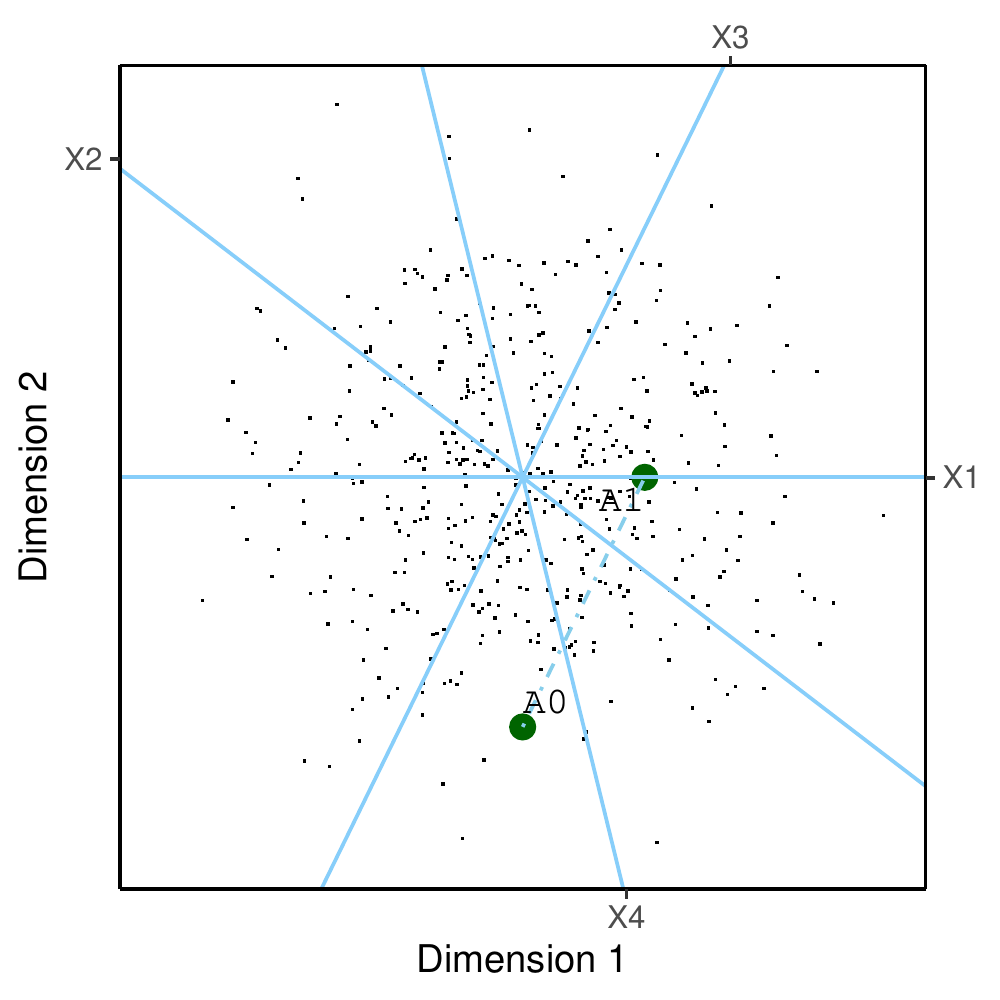}
\caption{Graphical representation with the two class points of a single response variable named A (the classes are represented by the points labeled A0 and A1), variable axes for four predictor variables, and subject points (small dots). The dashed-dotted line connecting A0 and A1 represents the vector $\mathbf{d}_A$. Predictor variable $X_3$ discriminates well between the two classes because the direction is parallel to the line joining the two classes, whereas predictor variable $X_2$ hardly discriminates between the two classes because the variable axis is almost orthogonal to the line joining the two classes.}
\label{fig:fig2}
\end{center}
\end{figure}

The variable axes can be understood as representations of subjects with
varying scores on the corresponding predictor variables and an average
score on all other predictor variables. In this way we can interpret the
variable axis by moving along the variable axes and computing the log
odds for each response variable. More formally, let us denote by
\(\mathbf{d}_r\) the \(M\)-vector with differences
\((v_{r1m} - v_{r0m})\) and let predictor variable $p$ be presented by its regression weights $\mathbf{b}_p$ (with $\mathbf{b}_p^\top$ row $p$ of matrix $\mathbf{B}$),
then we can write the effect of predictor
variable \(p\) on response variable \(r\) as 
\[
\mathbf{b}^\top_p\mathbf{d}_r = \|{\mathbf{b}_p}\| \cdot \|{\mathbf{d}_r}\| \cdot\cos(\mathbf{b}_p, \mathbf{d}_r),
\] 
showing that the log odds are largest when the direction of the variable
axis for predictor variable \(p\) is parallel to the line connecting the
two categories of response variable \(r\), while it is zero if the
variable axis is orthogonal to this line. Figure \ref{fig:fig2} illustrates this
property. In Figure \ref{fig:fig2}, the variable axes are
represented for four predictor variables. We use the convention that the
labels attached to the variable axes are placed on the positive side of the
variable. The two categories of a response variable (A0 and A1) are depicted by
points. The variable axis for predictor variable \(X_3\) is parallel to the
A01 line, indicating that this variable discriminates this response variable
well, while the variable axis for predictor \(X_2\) is almost orthogonal
to the A01 line, indicating that \(X_2\) does not discriminate between these two
classes. We could draw the projections of A0 and A1 on each of the
variable axes to see the discriminatory power: the further apart these
projections are, the higher the power. For example, the projections of the two
points onto variable \(X_2\) are very close to
each other, indicating that \(X_2\) does not discriminate these
two classes well. 

The discriminative power depends not only on the
distance between the projections but also on the estimated value of the
regression coefficients. Larger regression weights indicate more
general discriminative power for the complete set of response variables. 
We will indicate the value of the regression weight by using markers along the variable axis in steps of 1 standard
deviation. The further apart these markers are, the larger the
regression weights and the higher the discriminative power will be.

With \(R\) binary response variables, the number of different possible response
profiles is \(2^R\). When \(M = R\), each of these profiles can be perfectly
represented. In lower dimensional space ($M < R$), however, not all response
profiles find a place in the solution. For example, in a one-dimensional
space, only \(R+1\) different response profiles are represented. In two-dimensional space, the number of represented profiles is 
$$
\sum_{m=1}^2 {R \choose m} 
$$
\citep{coombs1955nonmetric}.

In the constrained models, the number of represented response profiles is
lower than in the general model because all decision lines are either
horizontal or vertical.\footnote{Assuming we do not have a response variable pertaining to multiple dimensions.}
With five response variables, of which three pertain to the first dimension
and two to the second, the model represents \(4 \times 3 = 12\) response
profiles, which is even smaller than the 16 in the general unconstrained model, and
much smaller than the \(2^R = 32\) possible response profiles.

\section{An MM Algorithm}

In this section, we develop an MM algorithm, where the first M stands for ``majorize'' and the second M for ``minimize''. Such algorithms are also known as 
iterative majorization (IM) algorithms. MM algorithms have the property of guaranteed descent and in MM algorithms it is easy to use low rank restrictions. 
The global idea of MM algorithms is that, instead of minimizing the original loss function, we seek an auxiliary function that 1) touches the original function at the current estimates, 2) lies above the original function, and 3) is easy to minimize. For a detailed treatment of the general principles of IM or MM, we refer to \cite{heiser1995convergent} and \cite{hunter2004tutorial}. In the following subsections, we will majorize a deviance function with a least squares function. This results in a fast algorithm. Before developing the algorithm, we will discuss identification of model parameters. 

%
%

\subsection{Admissible Transformations}\label{allowable-transformations}

Before we develop an algorithm for the estimation of model parameters, we must discuss indeterminacies, that is, admissible transformations that change neither the estimated probabilities nor the loss value. 

\begin{enumerate}
\item Multidimensional scaling and unfolding models in general have \emph{translational freedom}. We center $\mathbf{X}$ so that the origin of the Euclidean space is fixed at the average value of the predictor variables. 
\item The model has \emph{rotational freedom}: any map can be rotated without changing the distances or the probabilities. We will require that
\(\frac{1}{n}\mathbf{B}^\top\mathbf{X}^\top\mathbf{X}\mathbf{B} = \mathbf{I}\) so that the rotational indeterminacy is removed. Reflection can be removed by requiring that the regression weights for the first predictor variable are positive. 
\item The model 
$$
\pi_{rc}(\mathbf{x}_i) = \frac{\exp(\theta_{irc})}{\exp(\theta_{ir0}) + \exp(\theta_{ir1})}
$$
is invariant under an additive constant in the ``linear predictor'' ($\theta_{irc}$) , that is
\begin{eqnarray}
\frac{\exp(\theta_{irc})}{\exp(\theta_{ir0}) + \exp(\theta_{ir1})} = \frac{\exp(\theta_{irc} + \zeta_r)}{\exp(\theta_{ir0} + \zeta_r) + \exp(\theta_{ir1} + \zeta_r)}. 
\label{eq:indeterminacy}
\end{eqnarray}
For our distance model, $\theta_{irc} = -\delta(\mathbf{u}_i,\mathbf{v}_{rc})$, we can add a constant to the squared distances \textit{per response variable}, implying that the term $\sum_{m=1}^M u_{im}^2$ can be removed from the distance formulation. We will see a further simplification below. 
\end{enumerate}


\subsection{Algorithm for the unconstrained model}\label{sec:algorithm}

The deviance function to be minimized is
\begin{eqnarray}
L(\mathbf{B},\mathbf{V}) &=& -2\sum_{i=1}^n\sum_{r=1}^R\sum_{c=0}^{1}g_{irc} \log \pi_{rc}(\mathbf{x}_i) \\ \nonumber
&=& -2\sum_{i=1}^n\sum_{r=1}^R\sum_{c=0}^{1}g_{irc} \log \left(\frac{\exp (\theta_{irc})}{\exp(\theta_{ir0}) + \exp(\theta_{ir1})}\right),
\label{eq:deviance}
\end{eqnarray}
where for our model
\(\theta_{irc} = -\frac{1}{2}\delta(\mathbf{u}_i, \mathbf{v}_{rc})\). \cite{groenen2003}, \cite{deleeuw2006principal}, and 
\cite{groenen2016multinomial} show that the function \[
f_{ir}(\boldsymbol{\theta}_i) = -2\sum_{c=0}^{1}g_{irc} \log\frac{\exp (\theta_{irc})}{\exp(\theta_{ir0}) + \exp(\theta_{ir1})},
\] is majorized by \[
g_{ir}(\boldsymbol{\theta}_i,\tilde{\boldsymbol{\theta}}_i) = f_{ir}(\tilde{\boldsymbol{\theta}}_i) + (\boldsymbol{\theta}_i -  \tilde{\boldsymbol{\theta}}_i)^\top \nabla f_{ir}(\tilde{\boldsymbol{\theta}}_i) + \frac{1}{4} \|\boldsymbol{\theta}_i -  \tilde{\boldsymbol{\theta}}_i\|^2,
\] with \[
\nabla f_{ir}(\tilde{\boldsymbol{\theta}}_i) = -(\mathbf{g}_{ir} - \tilde{\boldsymbol{\pi}}_{ir}),
\] and where the tilde means that it is evaluated using the current
estimates. A proof of the majorizing property is given in the appendix of \cite{groenen2016multinomial}.

By analogy to Groenen and Josse, we therefore have that 
\begin{eqnarray}
L(\mathbf{B},\mathbf{V}) &\leq& \frac{1}{4}\left\|\mathbf{Z} - \boldsymbol{\Theta}\right\|^2 + L(\widetilde{\mathbf{B}}, \widetilde{\mathbf{V}}) -  \frac{1}{4}\left\|\mathbf{Z}\right\|^2  + \left\|\mathbf{G} - \widetilde{\boldsymbol{\Pi}}\right\|^2 
\nonumber \\
&=& \frac{1}{4}\left\|\mathbf{Z} - \boldsymbol{\Theta}\right\|^2 + \mbox{constant} = g(\mathbf{B},\mathbf{V}) + \mbox{constant},
\label{eq:majfunction}
\end{eqnarray} 
where \(\mathbf{Z} = \{z_{irc}\}\) with \(z_{irc} = \tilde{\theta}_{irc} + 2(g_{irc} - \tilde{\pi}_{irc})\) and
\(\boldsymbol{\Theta} = \{\theta_{irc}\}\) with \(\theta_{irc} = -\frac{1}{2}\sum_{m=1}^M (v_{rcm}^2 - 2u_{im}v_{rcm})\).
In matrix terms we can write \( \boldsymbol{\Theta} \) as \\
\[
\boldsymbol{\Theta} = -\frac{1}{2} \left({\mathbf{1}\mathbf{d}_{v}^\top - 2\mathbf{XB}\mathbf{V}^\top}\right),
\] 
and \(\mathbf{Z}\) as 
\[
\mathbf{Z} = \widetilde{\boldsymbol{\Theta}} + 2(\mathbf{G} - \widetilde{\boldsymbol{\Pi}}).
\]
Therefore, the objective function to be minimized in every iteration
is \[
g(\mathbf{B}, \mathbf{V}) = \left\|{ \mathbf{Z} + \mathbf{1}\mathbf{d}^{\top}_{v}/2- \mathbf{XB}\mathbf{V}^{\top} }\right\|^2.
\]

The third indeterminacy, outlined above, allows us to rewrite the minimization function as 
\begin{eqnarray}
g(\mathbf{B}, \mathbf{V}) = \left\|{\mathbf{ZJ} + \frac{1}{2}\mathbf{1}\mathbf{d}^{\top}_{v}\mathbf{J}- \mathbf{XB}\mathbf{V}^{\top}\mathbf{J}}\right\|^2,
\label{eq:majfunction2}
\end{eqnarray}
with \(\mathbf{J}\) a symmetric block diagonal matrix with \(2 \times 2\) diagonal blocks \(\mathbf{J}_r\), the usual centering matrices,
\(\mathbf{J}_r = \mathbf{I}_2 - \frac{1}{2}\mathbf{11}^\top\).

Let us define the matrices
\(\mathbf{A}_l = \mathbf{I}_R \otimes [1, 1]^\top\) and
\(\mathbf{A}_k = \mathbf{I}_R \otimes [1, -1]^\top\) with \(\otimes\) the
Kronecker product, such that \(\mathbf{V}\) can be reparametrized as \[
\mathbf{V} = \mathbf{A}_l \mathbf{L} + \mathbf{A}_k \mathbf{K}
\] with \(\mathbf{L}\) the \(R \times M\) matrix with response variable
\textit{locations} and \(\mathbf{K}\) the \(R \times M\) matrix representing the
\textit{discriminatory power} for the response variables. As a numerical example with two response variables, consider 
$$
\mathbf{V} = \left[\begin{array}{cc}
1 & 0 \\
3 & 2 \\ \hdashline
0 & 0 \\
4 & 6
\end{array}\right]
= 
\left[\begin{array}{cc}
1 & 0 \\
1 & 0 \\
0 & 1 \\
0 & 1
\end{array}\right] 
\left[\begin{array}{cc}
2 & 1 \\
2 & 3 \\
\end{array}\right] 
+ 
\left[\begin{array}{rr}
1 & 0 \\
-1 & 0 \\
0 & 1 \\
0 & -1
\end{array}\right] 
\left[\begin{array}{rr}
-1 & -1 \\
-2 & -3 \\
\end{array}\right], 
$$
where the second matrix on the right-hand side of the equation represents $\mathbf{L}$ (the locations or midpoints of the class coordinates) and the last matrix on the right-hand side of the equation represents $\mathbf{K}$. The larger the absolute values in row $r$ of matrix $\mathbf{K}$,
the better the two categories from this response variable ($r$) can be
discriminated by the predictor variables; if the values equal zero, the
categories cannot be discriminated and the positions of the two
categories of a response variable fall in the same place. In the numerical example, the second response variable is better discriminated (that is, the class points are further apart). 

Using this reparametrization in our loss function, the last term \(\mathbf{XB}\mathbf{V}^{\top}\mathbf{J}\) simplifies to
\(\mathbf{XB}\mathbf{V}^{\top}\mathbf{J} = \mathbf{XB}\mathbf{K}^{\top}\mathbf{A}^\top_k\) because the term 
$\mathbf{XB}\mathbf{L}^{\top}\mathbf{A}^\top_l \mathbf{J} = \mathbf{0}$ as $ \mathbf{A}^\top_l \mathbf{J} = \mathbf{0}$. Let us now have a closer look at the second term of the loss function (Equation \ref{eq:majfunction2}), that is, 
\(\mathbf{1}\mathbf{d}_{v}^\top\mathbf{J} = \mathbf{J}\mathbf{d}_{v}\). This term can be rewritten as
\begin{eqnarray*}
\mathbf{d}_{v} = \mathrm{Diag}(\mathbf{VV}^\top) &=& \mathrm{Diag} \left( (\mathbf{A}_l \mathbf{L} + \mathbf{A}_k \mathbf{K})(\mathbf{A}_l \mathbf{L} + \mathbf{A}_k \mathbf{K})^\top \right)\\
 &=& \mathrm{Diag} \left( \mathbf{A}_l\mathbf{LL}^\top\mathbf{A}_l^\top + \mathbf{A}_k\mathbf{KK}^\top\mathbf{A}_k^\top + \mathbf{A}_l\mathbf{LK}^\top\mathbf{A}_k^\top + \mathbf{A}_k\mathbf{KL}^\top\mathbf{A}_l^\top \right),
\end{eqnarray*} 
where $\mathrm{Diag}(\mathbf{X})$ creates a column vector of the diagonal elements of $\mathbf{X}$. 
It can be verified that the terms
\(\mathbf{J}\mathrm{Diag} \left( \mathbf{A}_l\mathbf{LL}^\top\mathbf{A}_l^\top \right)\)
and
\(\mathbf{J}\mathrm{Diag}\left( \mathbf{A}_k\mathbf{KK}^\top\mathbf{A}_k^\top \right)\)
are both equal to zero. Therefore  
\[
\mathrm{Diag}(\mathbf{VV}^\top) = \mathbf{J} \mathrm{Diag} \left(2 \mathbf{A}_k\mathbf{KL}^\top\mathbf{A}_l^\top \right).
\] 
Focusing on a single response variable \(r\), we can rewrite the corresponding part of the previous equation as 
\[
\mathbf{k}^\top_r \mathbf{l}_r \left[ \begin{array}{rr} 1 & 1\\ -1 & -1 \end{array}\right],
\]
from which it follows that \(\frac{1}{2}\mathbf{1}\mathbf{d}_{v}\mathbf{J}\) can be
written as 
\[
\frac{1}{2}\mathbf{1}\mathbf{d}_{v}^\top\mathbf{J} = \mathbf{1}\mathrm{Diag}(\mathbf{KL}^\top)^\top\mathbf{A}_k^\top.
\] 

Going back to our loss function and using the decomposition into \(\mathbf{K}\) and \(\mathbf{L}\), it becomes 
\[
g(\mathbf{B}, \mathbf{K}, \mathbf{L}) = \left\|{\mathbf{ZJ} + \mathbf{1}\mathrm{Diag}(\mathbf{KL}^\top)^\top\mathbf{A}_k^\top- \mathbf{XB}\mathbf{K}^{\top}\mathbf{A}^\top_k}\right\|^2,
\] 
that equals 
\[
g(\mathbf{B}, \mathbf{K}, \mathbf{L}) = 2 \left\|{\frac{1}{2}\mathbf{ZJA}_k + \mathbf{1}\mathrm{Diag}(\mathbf{KL}^\top)^\top- \mathbf{XB}\mathbf{K}^{\top}}\right\|^2 
\] 
and can be rewritten as 
\[
g(\mathbf{B}, \mathbf{K}, \mathbf{L}) = 2\left\|{\frac{1}{2}\mathbf{ZJA}_k + \mathbf{1}\mathbf{a}^\top- \mathbf{XB}\mathbf{K}^{\top}}\right\|^2,
\] 
where $\mathbf{a}$ is a vector with elements $a_r = \sum_{m=1}^M k_{rm}l_{rm}$. The elements of $\mathbf{a}$ can be estimated independently of $\mathbf{K}$, because values of $\mathbf{L}$ always exist that together with the $k_{rm}$ can reconstruct $a_r$ (see below). Therefore, this latter loss function can be solved separately for 1) \(\mathbf{a}\) and 2) \(\mathbf{B}, \mathbf{K}\).

To update  \(\mathbf{a}\), define
\[
\tilde{\mathbf{Z}}_1 = \frac{1}{2}\mathbf{ZJA}_k,
\] 
then the update is 
\(\mathbf{a}^{+} = -\tilde{\mathbf{Z}}_1^\top\mathbf{1}/n\). 

To find the update for \(\mathbf{B}\) and \(\mathbf{K}\), let us define 
\[
\tilde{\mathbf{Z}}_2 = \frac{1}{2}\mathbf{ZJA}_k + \mathbf{1a}^\top,
\] 
so that we have to minimize
\[
\left\| \tilde{\mathbf{Z}}_2- \mathbf{XB}\mathbf{K}^{\top}\right\|^2,
\]
under the restriction that \(n^{-1}\mathbf{B}^\top\mathbf{X}^\top\mathbf{X}\mathbf{B} = \mathbf{I}\). 
As 
\[\left\|{\tilde{\mathbf{Z}}_2- \mathbf{XB}\mathbf{K}^{\top}}\right\|^2 = \left\| \tilde{\mathbf{Z}}_2- \mathbf{XN} \right\|^2 + \left\| \mathbf{N} - \mathbf{BK}^{\top}\right\|_{\mathbf{X}^\top\mathbf{X}}^2
\]
with 
$\mathbf{N} = (\mathbf{X}^\top\mathbf{X})^{-1}\mathbf{X}^\top\tilde{\mathbf{Z}}_2$, the unconstrained update, an update of \(\mathbf{B}\) and \(\mathbf{K}\) is found by the generalized singular value decomposition of $\mathbf{N}$ \citep{takane2013constrained}. These two steps can be combined, that is,
\[
\mathbf{R}^{-1}_x\mathbf{X}^{\top} \tilde{\mathbf{Z}}_2 = \mathbf{P}\boldsymbol{\Phi}\mathbf{Q}^\top,
\] 
where \(\mathbf{R}_x\) is the matrix square root of the matrix \(\mathbf{X}^\top\mathbf{X}\),
that is~\(\mathbf{X}^\top\mathbf{X} = \mathbf{R}_x\mathbf{R}_x^\top\). The
updates for \(\mathbf{B}\) and \(\mathbf{K}\) can be obtained as 
\[
\mathbf{B}^{+} = \sqrt{n} \mathbf{R}_x^{-1}\mathbf{P}_M,
\] 
where \(\mathbf{P}_M\) denotes the \(M\) columns of \(\mathbf{P}\)
corresponding to the $M$ largest singular values and 
\[
\mathbf{K}^{+} = \frac{1}{\sqrt{n}}\mathbf{Q}_M\boldsymbol{\Phi}_M.
\] 
Finally, an update for \(\mathbf{L}\) can be obtained from
\(\mathbf{a}\) and \(\mathbf{K}\). For every response variable we have
\[
a_r = \sum_{m=1}^M k_{rm}l_{rm},
\] 
which is an unidentified system, that is, there are many choices of
\(l_{rm}\) that provide a solution. These solutions correspond to any position on the decision line (or plane or hyperplane in higher-dimensional spaces), as discussed in relation to Figure \ref{fig:fig1}. We find the position on this hyperplane that is closest to the origin of the Euclidean space, that is,
$$
l_{rm}^+ = \frac{a_r k_{rm}}{\sum_m k_{rm}^2}.
$$
A summary of the algorithm can be found in Algorithm \ref{melodic.alg}. 


The number of parameters of the model is:
\begin{itemize}
\item
  \(PM - M(M+1)/2\) for the regression weights \(\mathbf{B}\);
\item
  \(RM\) parameters in the matrix \(\mathbf{K}\);
\item
  \(R\) parameters in \(\mathbf{L}\);
\end{itemize}
which sum to \((P + R)M + R - M(M+1)/2\).

\begin{algorithm}[H]
 \SetAlgoLined 
 \KwData{$\mathbf{X}$, $\mathbf{G}$, $M$}
 \KwResult{$\mathbf{B}$, $\mathbf{K}$, $\mathbf{L}$}
 Compute: $\mathbf{R}_x^{-1}\mathbf{X}^{\top} \mathbf{G} = \mathbf{P}\boldsymbol{\Phi}\mathbf{Q}^\top$\;
 Initialize: $\mathbf{B}^{(0)} = \sqrt{n} \mathbf{R}_x^{-1}\mathbf{P}_M$ \;
 Compute: $\mathbf{V} = \frac{1}{\sqrt{n}}\mathbf{Q}_M\boldsymbol{\Phi}_M$\;
 Initialize: $\mathbf{K}^{(0)}$ by taking the uneven rows of $\mathbf{JV}$\;
 Initialize: $\mathbf{L}^{(0)}$ by taking the uneven rows of $(\mathbf{I} - \mathbf{J})\mathbf{V}$\;
 Compute: $\tilde{\boldsymbol{\Pi}}$ and $\tilde{\boldsymbol{\Theta}}$\;
 \While{$t = 0$ or $(L^t - L^{(t-1)})/L^t  > 10^{-8}$}{
  $t = t + 1$\;
  Compute: $\mathbf{Z} = \tilde{\boldsymbol{\Theta}} + 2(\mathbf{G} - \tilde{\boldsymbol{\Pi}}) $\;
  Compute: $ \mathbf{Z}_1 = \frac{1}{2}\mathbf{ZJA}_k$ \;
  Compute: $\mathbf{a} = - \mathbf{Z}_1^\top\mathbf{1}/n$ \;
  Compute: $\mathbf{Z}_2 = \frac{1}{2}\mathbf{ZJA}_k + \mathbf{1a}^\top$\;
  Compute: $\mathbf{R}_x^{-1}\mathbf{X}^{\top} \mathbf{Z}_2 = \mathbf{P}\boldsymbol{\Phi}\mathbf{Q}^\top$\;
  Update: $\mathbf{B}^{(t)} = \sqrt{n} \mathbf{R}_x^{-1}\mathbf{P}_M$ \;
  Update: $\mathbf{K}^{(t)} = \frac{1}{\sqrt{n}}\mathbf{Q}_M\boldsymbol{\Phi}_M$\;
  Update: $l_{rm}^{(t)} = a_r k_{rm}/(\sum_m k_{rm}^2), \forall r$\;
  Compute $\tilde{\boldsymbol{\Pi}}$, $\tilde{\boldsymbol{\Theta}}$, and $L^t$\;
 }
 \caption{MELODIC Algorithm \label{melodic.alg}}
\end{algorithm}

\subsection{Algorithm for constrained model}\label{constrained-model.}

Sometimes researchers have an idea  in advance of which responses belong
together in which dimensions. Let us denote the set of response variables that pertains to dimension $m$ by
$\mathcal{D}_m$. Furthermore, let us denote the set of dimensions to which response variable $r$ pertains as $\mathcal{S}_r$.
Then, for $m \notin \mathcal{S}_r$, we restrict $l_{rm} = k_{rm} = 0$.

Much of the unconstrained MM algorithm can be used, except for two aspects: (i) de orthonormality restriction on $\mathbf{XB}$ needs to be relaxed and (2) the updates will be done dimension wise. 
We still need a scale restriction per dimension on $\mathbf{XB}$. The equivalent of (\ref{eq:majfunction2}) can be written 
\[
g_c(\mathbf{B}, \mathbf{K}, \mathbf{L}) = \left\|{\frac{1}{2}\mathbf{ZJA}_k + \mathbf{1}\mathbf{a}^\top- \sum_{m=1}^M \mathbf{Xb}_m\mathbf{k}_m^{\top}}\right\|^2,
\] 
which shows that the last term can be decomposed in dimensional
terms. To update for dimension \(s\), we first define
\[
\tilde{\mathbf{Z}}_3 = \frac{1}{2}\mathbf{ZJA}_k + \mathbf{1a}^\top - \sum_{m \neq s}\mathbf{Xb}_m\mathbf{k}_m^{\top}.
\] 
Next, we define the matrix $\tilde{\mathbf{Z}}_s$ to be the subset of the matrix $\tilde{\mathbf{Z}}_3$ consisting of the columns for which $r \in \mathcal{D}_s$. 

Similar to the unconstrained model, a generalized singular value decomposition of \(\tilde{\mathbf{Z}}_s\) gives updates for $\mathbf{b}_s$ and $\mathbf{k}_s$ by taking the highest singular value and the corresponding vector. The update for \(\mathbf{a}\) is
the same as in the unconstrained model. The update for \(\mathbf{L}\) is
similar to the unconstrained model, but we only give non-zero values to
the dimensions to which response variable \(r\) pertains. A summary of the algorithm is given in Algorithm \ref{melodic.alg2}.


The total number of parameters for the constrained model depends on the
specific constraints. Nevertheless, we have
\begin{itemize}
\item
  \((P-1)M\) parameters for the regression weights \(\mathbf{B}\);
\item
  We define an indicator matrix of size \(R \times M\) indicating which
  response belongs to which dimension. The number of parameters in the
  matrix \(\mathbf{K}\) equals the number of ones in that indicator
  matrix (see the next section for an example);
\item
  \(R\) parameters in \(\mathbf{L}\).
\end{itemize}


\begin{algorithm}[H]
 \SetAlgoLined 
 \KwData{$\mathbf{X}$, $\mathbf{G}$, $M$}
 \KwResult{$\mathbf{B}$, $\mathbf{K}$, $\mathbf{L}$}
 Compute: $\mathbf{R}_x^{-1}\mathbf{X}^{\top} \mathbf{G} = \mathbf{P}\boldsymbol{\Phi}\mathbf{Q}^\top$\;
 Initialize: $\mathbf{B}^{(0)} = \sqrt{n} \mathbf{R}_x^{-1}\mathbf{P}_M$ \;
 Compute: $\mathbf{V} = \frac{1}{\sqrt{n}}\mathbf{Q}_M\boldsymbol{\Phi}_M$\;
 Initialize: $\mathbf{K}^{(0)}$ by taking the uneven rows of $\mathbf{JV}$\;
 Initialize: $\mathbf{L}^{(0)}$ by taking the uneven rows of $(\mathbf{I} - \mathbf{J})\mathbf{V}$\;
 Set elements in $\mathbf{K}^{(0)}$ and $\mathbf{L}^{(0)}$ to zero, following the constraints\;
 Compute: $\tilde{\boldsymbol{\Pi}}$ and $\tilde{\boldsymbol{\Theta}}$\;
 \While{$t = 0$ or $(L^t - L^{(t-1)})/L^t  > 10^{-8}$}{
  $t = t + 1$ \;
  Compute: $\mathbf{Z} = \tilde{\boldsymbol{\Theta}} + 2(\mathbf{G} - \tilde{\boldsymbol{\Pi}}) $\;
  Compute: $\tilde{\mathbf{Z}}_1 = \frac{1}{2}\mathbf{ZJA}_k$ \;
  Compute: $\mathbf{a}^+ = -\tilde{\mathbf{Z}}_1^\top\mathbf{1}/n$ \;
  \For{$s = 1,\ldots,M$}{
  Compute: $\tilde{\mathbf{Z}}_s$\;
  Compute: $\mathbf{R}_x^{-1}\mathbf{X}^{\top} \tilde{\mathbf{Z}}_s = \mathbf{P}\boldsymbol{\Phi}\mathbf{Q}^\top$\;
  Update: $\mathbf{b}_s^{(t)} = \sqrt{n} \mathbf{R}_x^{-1}\mathbf{P}_1$ \;
  Update: $\mathbf{k}_s^{(t)} = \frac{1}{\sqrt{n}}\mathbf{Q}_1\boldsymbol{\Phi}_1$\;
  }
  Update: $l_{rm}^{(t)} = a_r/\sum_{m=1}^M k_{rm}$, for $m \in \mathcal{S}_r$ and $\forall r$\;
  Compute $\tilde{\boldsymbol{\Pi}}$, $\tilde{\boldsymbol{\Theta}}$, and $L^t$\;
 }
 \caption{MELODIC Algorithm for Constrained Model \label{melodic.alg2}}
\end{algorithm}

\section{Two empirical applications}

In this section we discuss two empirical applications of the model. The first data set considers profiles of drug consumption; the second, profiles of mental disorders. For the first data set we use the unconstrained model. For the second data set we start with a set of constrained models representing different theories. 

\subsection{Drug Consumption Data}

The drug consumption data \citep{fehrman2017five} has records for 1885 respondents. For each
respondent, nine attributes are measured. We have personality measurements based on the big five personality traits, neuroticism (N), extraversion (E), openness to experience (O),
agreeableness (A), and conscientiousness (C), and two other personality characteristics, namely impulsivity (I) and sensation seeking (S). Data were also collected on age and gender.\footnote{Also level of education, ethnicity, and country of origin are available in the original data base. We omitted these from the analysis.}

In addition, participants were questioned concerning their use of 18
legal and illegal drugs. For each drug, participants were asked whether they never used the drug, used it over a decade ago, in
the last decade, in the last year, month, week, or day. In our analysis
we coded whether participants used the particular drug in the last year (yes or no).
Furthermore, in our analysis we focused on the drugs that had a minimum percentage of
10\% and a maximum of 90\%, which are Amphetamine, Benzodiazepine,
Cannabis, Cocaine, Ecstasy, Ketamine, legal highs, LSD,
Methadone, Mushrooms, and Nicotine (\(R = 11\)).


\begin{table}
\centering
\begin{tabular}{crrrr}
  \hline
Dimensionality & Deviance & \#param & AIC & BIC \\ 
  \hline
1 & 18311 & 30  & 18371 & 18538 \\ 
2 & 18117 & 48  & 18213 & {\bf 18479} \\ 
3 & 18030 & 65  & {\bf 18160} & 18520 \\ 
4 & 17998 & 81  & 18160 & 18609 \\ 
5 & 17987 & 96  & 18179 & 18711 \\ 
6 & 17980 & 110 & 18200 & 18810 \\ 
7 & 17975 & 123 & 18221 & 18903 \\ 
   \hline
\end{tabular}
\caption{AIC and BIC statistics for models in 1 to 7 dimensions for the drug consumption data.}
\label{tab:AICdim}
\end{table}

The first step in the analysis is to select the dimensionality. We fit
models in one to seven dimensions and compute information criteria statistics for
comparison. The results are given in Table \ref{tab:AICdim}, where we can see that either the two- or three-dimensional solution is
optimal according to the AIC and BIC statistics. 

We should also check the influence of the predictor variables. Each of the predictor variables is left out of the two-dimensional model. The AIC and BIC statistics are shown in Table \ref{tab:AICpred}, where it can be seen that only impulsivity can be considered for being left out of the model; all other predictors would lead to a substantial loss in fit. We decided to further interpret the model using all predictor variables except impulsivity. 

\begin{table}
\centering
\begin{tabular}{lrrrr}
  \hline
 Left Out & Deviance & \#param & AIC & BIC \\ 
  \hline
 Age 				& 19303 & 46 & 19395 & 19650 \\ 
 Gender 			& 18417 & 46 & 18509 & 18764 \\ 
 Neuroticism 		& 18181 & 46 & 18273 & 18528 \\ 
 Extraversion 		& 18134 & 46 & 18226 & 18481 \\ 
 Openess 			& 18449 & 46 & 18541 & 18796 \\ 
 Agreeablenes 		& 18137 & 46 & 18229 & 18484 \\ 
 Conscientiousness 	& 18172 & 46 & 18264 & 18519 \\ 
 Impulsivity 		& 18121 & 46 & 18213 & {\bf 18468} \\ 
 Sensation seeking 	& 18409 & 46 & 18501 & 18756 \\ 
   \hline
\end{tabular}
\caption{AIC and BIC statistics for two-dimensional models with 1 predictor left out of the model.}
\label{tab:AICpred}
\end{table}

\begin{sidewaysfigure}
\begin{center}
\includegraphics[width = .95\textwidth]{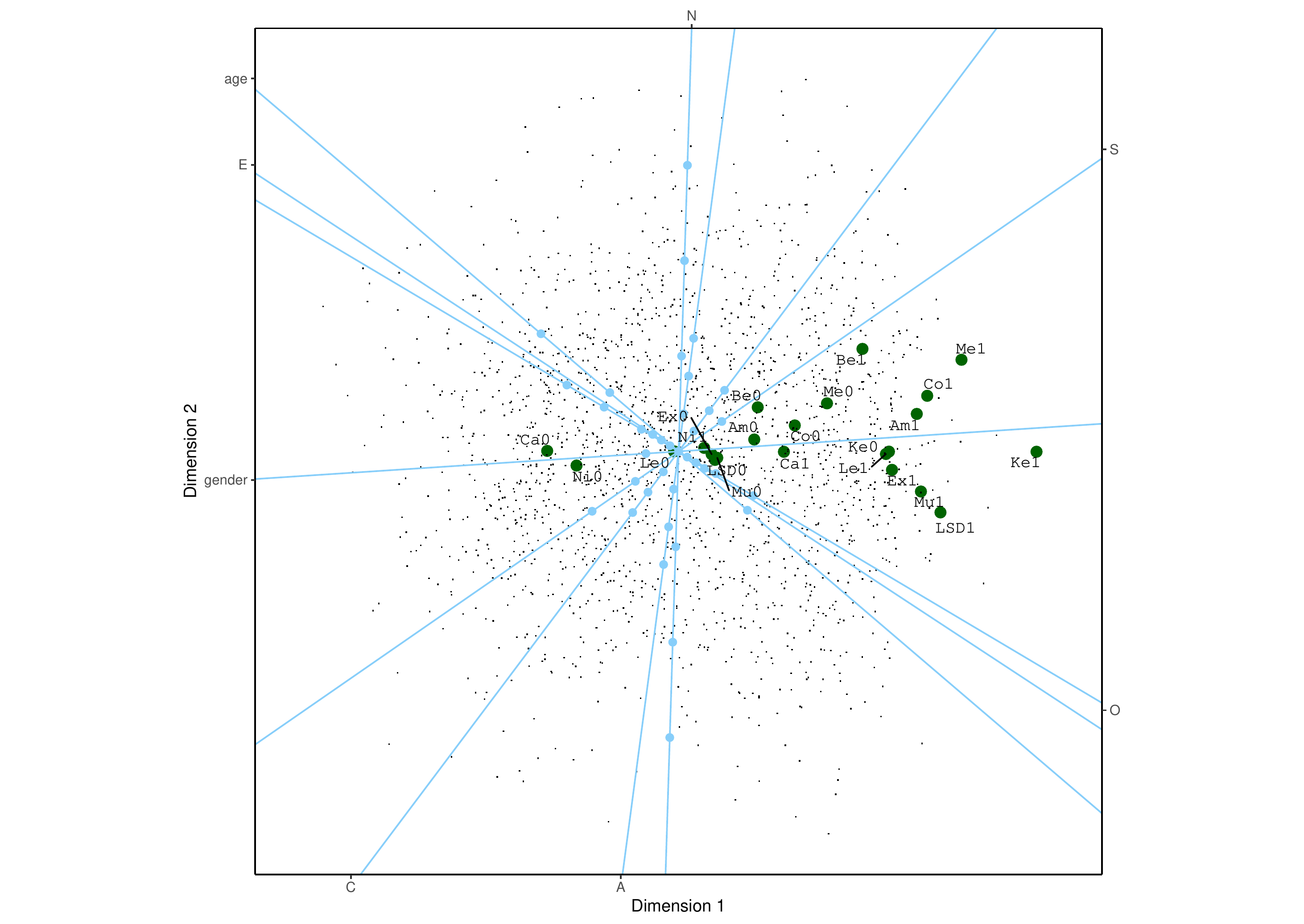}
\caption{Two-dimensional solution for the drug consumption data. Predictor variable labels are printed on the border of the graph where N = Neuroticism; E = Extraversion; O = Openness to experience; A = Agreeableness; C = Conscientiousness; S = Sensation seeking. We use the convention that labels are printed at the positive side of the variable. Markers on the variable axes indicate standard deviation increases/decreases from the mean. Category points are labeled by the name of the drug together with a 1 (yes) or 0 (no), that is, \texttt{Am1} denotes the class point for use of Amphetamine and \texttt{Am0} for no use of Amphetamine.  Am = Amphetamine; Be = Benzodiazepine; Ca = Cannabis; Co = Cocaine; Ex =  Ecstasy; Ke = Ketamine; Le =  legal highs; LSD = LSD; Me = Methadone; Mu = Mushrooms; Ni = Nicotine.}
\label{fig:drug1}
\end{center}
\end{sidewaysfigure}

The graphical representation of the two-dimensional model is shown in Figure \ref{fig:drug1}.\footnote{We left out the decision lines to avoid clutter.}
The first thing that catches the eye in Figure \ref{fig:drug1} is that the categories for "yes" (labels ending with 1) are all to the right-hand side of the categories for "no" (labels ending with 0). Therefore, participants who use drugs lie on the positive side of the first dimension. It can be seen that cannabis is furthest to the left, while Ketamine, LSD and Methadone are on the extreme right-hand side. Apparently, if participants start using drugs, they start with cannabis and only add other drugs later. 

Considering the predictor side, we see that five predictor variables run from left to right: openness to experience (O), sensation seeking (S), extraversion (E), age, and gender. With respect to gender and age, boys use more drugs than girls and younger people use more drugs than the elderly. Furthermore, participants who are more open to experience (O) and less extravert (E) use more drugs. Finally, participants scoring high on sensation seeking (S) use more drugs than participants who score low on sensation seeking. 

The vertical dimension is harder to interpret because the differences between the yes and no points are often small. The largest differences are for benzodiazepine and methadone (yes category has a higher coordinate) and for LSD (yes category has a lower coordinate). The predictor variable neuroticism (N) points strongly in this direction, indicating that neurotic participants tend to use  benzodiazepine and methadone more frequently but LSD less frequently. The variable axis for agreeableness (A) is almost parallel to the vertical dimension, but in the opposite direction to neuroticism, indicating opposite effects.  

To get a more detailed interpretation, let us look at the estimated implied logistic regression coefficients (Equation \ref{eq:lrcoefs}) in Table \ref{tab:drugcoef}). Since the predictor variables are standardized to have zero mean and standard deviation one, these are changes in log odds for one standard deviation increases in the predictors. The numbers in each column can be interpreted as the standardized coefficients in a single logistic regression model.

\begin{table}[t]
\centering
\begin{tabular}{l|rrrrrrrrrrr}
  \hline
  & \multicolumn{11}{c}{Response Variables} \\
 Predictor variables & Am & Be & Ca & Co & Ex & Ke & Le & LSD & Me & Mu & Ni \\ 
  \hline
age & -0.59 & -0.23 & -0.99 & -0.45 & -0.81 & -0.62 & -0.90 & -1.13 & -0.41 & -0.97 & -0.47 \\ 
  gender & -0.33 & -0.22 & -0.47 & -0.27 & -0.36 & -0.29 & -0.42 & -0.44 & -0.27 & -0.40 & -0.26 \\ 
  Neuroticism & 0.18 & 0.36 & 0.04 & 0.20 & -0.06 & 0.03 & 0.02 & -0.26 & 0.28 & -0.16 & 0.13 \\ 
  Extraversion & -0.08 & -0.03 & -0.12 & -0.06 & -0.10 & -0.08 & -0.11 & -0.14 & -0.06 & -0.12 & -0.06 \\ 
  Openess & 0.33 & 0.16 & 0.54 & 0.26 & 0.43 & 0.33 & 0.48 & 0.58 & 0.25 & 0.51 & 0.27 \\ 
  Agreeableness & -0.11 & -0.17 & -0.07 & -0.11 & -0.02 & -0.04 & -0.06 & 0.05 & -0.14 & 0.02 & -0.08 \\ 
  Conscientiousness & -0.18 & -0.17 & -0.22 & -0.16 & -0.15 & -0.14 & -0.19 & -0.14 & -0.18 & -0.15 & -0.14 \\ 
  Sensation seeking & 0.47 & 0.38 & 0.62 & 0.40 & 0.45 & 0.39 & 0.55 & 0.50 & 0.43 & 0.47 & 0.37 \\ \hline
  Quality & 1.00 & 0.96 & 0.97 & 0.90 & 0.96 & 0.95 & 1.00 & 0.98 & 0.94 & 0.99 & 0.98 \\
   \hline
 \end{tabular}
\caption{Estimated implied regression coefficients (equation \ref{eq:lrcoefs}) of each of the predictor variables for each of the response variables. 
In the columns Am = Amphetamine; Be = Benzodiazepine; Ca =Cannabis; Co = Cocaine; Ex =  Ecstasy; Ke = Ketamine; LE =  legal highs; 
LSD = LSD; Me = Methadone; Mu = Mushrooms; Ni = Nicotine. Last line shows the quality of representation ($Q_r$). }
\label{tab:drugcoef}
\end{table}

We can verify how well each response variable is represented in the low-dimensional space. To do this, we define a measure called Quality of
Representation, \(Q_r\), which is defined by 
\[
Q_r = (L_{(0,r)} - L_r)/(L_{(0,r)} - L_{lr}), 
\] 
where \(L_{(0,r)}\) is the deviance of the intercept-only logistic
regression model for response variable \(r\), \(L_r\) is the part of our
loss function for response variable $r$, and \(L_{lr}\) is the deviance from a
logistic regression with the same predictor variables. Thus, $Q_r $can be interpreted as the proportion of loss in deviance imposed by the Melodic model compared to an unconstrained logistic regression for response variable $r$. The quality of representation for the response variables in this analysis are given in the last row of Table \ref{tab:drugcoef}, where it can be seen that most response variables are well represented. The response variable ``cocaine'' (Co) is worst represented, although still with 89.8\% recovered. 


\subsection{Depression and Anxiety data}

Depression and anxiety disorders are common at all ages. Approximately
one out of three people in the Netherlands will be faced with them at
some time during their lives. It is not clear why some people
recover quickly and why others suffer for long periods of time. The
Netherlands Study of Depression and Anxiety (NESDA) was therefore
designed to investigate the course of depression and anxiety disorders
over a period of several years. For more information about the study
design, see \cite{penninx2008netherlands}. In our application, we will analyze
data from the first wave, focusing on the relationship between
personality and depression and anxiety disorders. The data were
previously analyzed by \cite{spinhoven2009role}. Data were collected from
three different populations: from primary health care; from generalized
mental health care; and from the general population. Our analysis will
focus on the population of generalized health care.

We have data for 786 participants. The diagnoses Dysthymia (D), Major Depressive Disorder (MDD), 
Generalized Anxiety Disorder (GAD), Social Phobia (SP), and Panic
Disorder (PD) were established with the Composite Interview
Diagnostic Instrument (CIDI) psychiatric interview. Personality was
operationalized using the 60-item NEO Five-Factor Inventory (NEO-FFI).
The NEO-FFI questionnaire measures the following five personality
domains: Neuroticism, Extraversion, Agreeableness, Conscientiousness and
Openness to Experience. In addition to these five predictors, three background
variables were measured: age, gender, and education in years.

The prevalences in the data are
21.25\% for dysthymia, 76.21\% for major depressive disorder, 30.41\%
for generalized anxiety disorder, 41.6\% for social phobia, and 52.8\%
for panic disorder. Of the 786 participants, 272 have a single disorder,
the others all have multiple disorders. There are 235 participants with
two disorders, 147 with three, 96 with four, and 36 participants with five
disorders.

Due to the high comorbidity among disorders, the scientific field of psychiatry developed
three different theories:
\begin{enumerate}
\item a unidimensional structure where all the disorders are represented by a single dimension;
\item a two-dimensional structure with one dimension representing distress (D, MDD, GAD) and the other fear (SP, PD);
\item a two-dimensional structure with one dimension representing depression (D, MDD) and the other anxiety (GAD, SP, PD).
\end{enumerate}
We can of course define another two-dimensional structure (Theory 4) in which
dysthymia and major depressive disorder pertain to the first
dimension, social phobia and panic disorder to the second, and generalized anxiety disorder to both dimensions.

Each of the three two-dimensional theories gives rise to a different response variable by
dimension indicator matrix: \[
\mathbf{D}_2 = \left[\begin{array}{cc} 
1 & 0\\
1 & 0\\ 
1 & 0\\
0 & 1\\
0 & 1\end{array}\right],   
\ 
\mathbf{D}_3 = \left[\begin{array}{cc} 
1 & 0\\
1 & 0\\ 
0 & 1\\
0 & 1\\
0 & 1\end{array}\right], 
\ 
\mathbf{D}_4 = \left[\begin{array}{cc} 
1 & 0\\
1 & 0\\ 
1 & 1\\
0 & 1\\
0 & 1\end{array}\right]. 
\]

We fitted the four models reflecting the four theories to the data. The fit statistics can be found in Table \ref{tab:aic.nesda}, where it can be seen that all four theories give about the same fit, but the distress-fear hypothesis (corresponding to $\mathbf{D}_2$) has a slightly lower AIC and the unidimensional model a lower BIC value than the other theories. 

\begin{table}[ht]
\centering
\begin{tabular}{crrrrr}
  \hline
Theory/Model & Deviance & \#param & AIC & BIC \\ 
  \hline
  1 & 4553.34 & 17 & 4587 & 4667 \\ 
  2 & 4531.17 & 24 & 4579 & 4691 \\ 
  3 & 4533.71 & 24 & 4582 & 4694 \\ 
  4 & 4530.35 & 25 & 4580 & 4697 \\ 
   \hline
\end{tabular}
\caption{AIC and BIC statistics for the models reflecting the four theories.}
\label{tab:aic.nesda}
\end{table}

\begin{sidewaysfigure}
\begin{center}
\includegraphics[width = .95\textwidth]{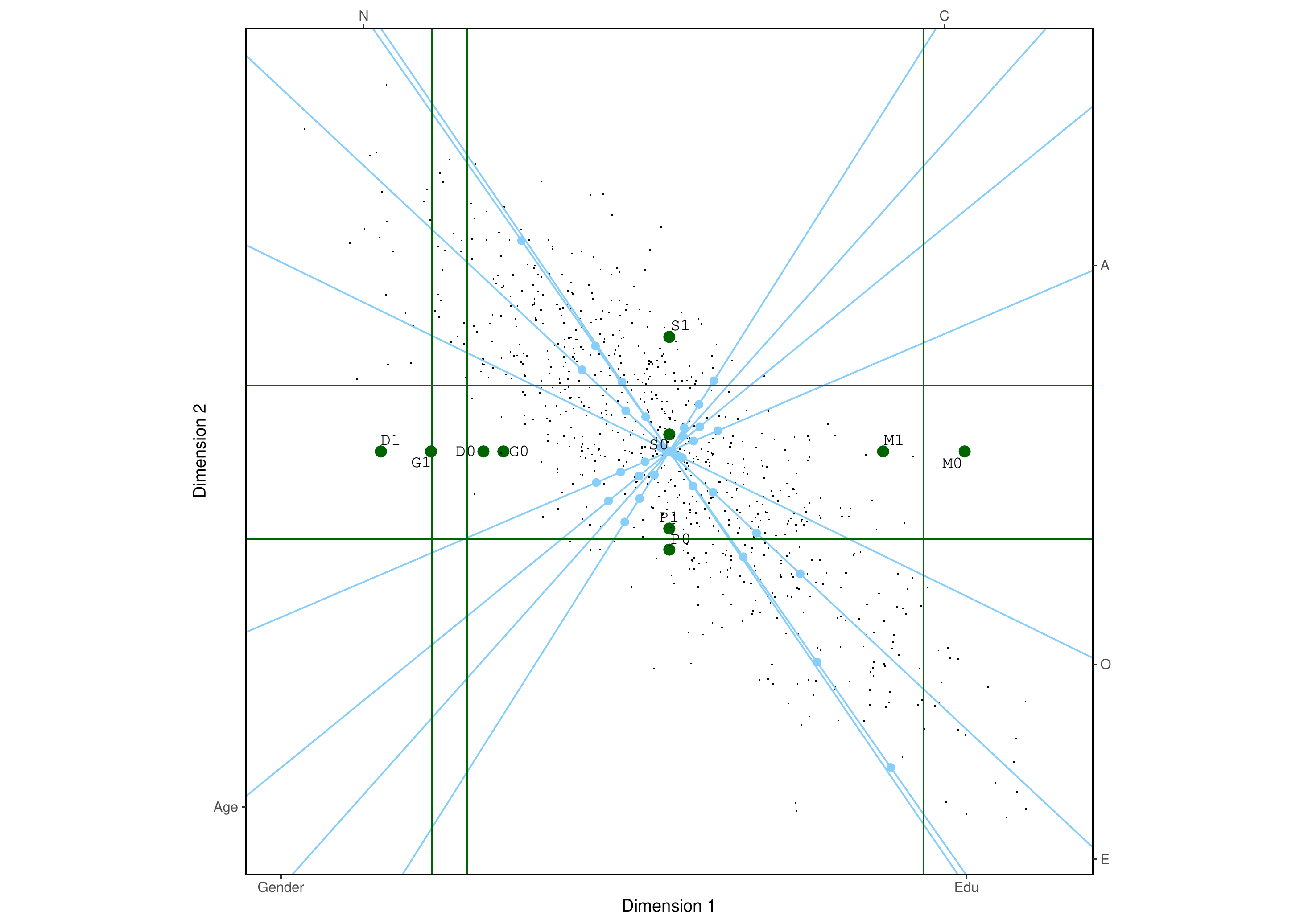}
\caption{Two-dimensional solution following Theory 2 for the depression and anxiety data. Predictor variable labels are printed on the border of the graph where N = Neuroticism; E = Extraversion; O = Openness to experience; A = Agreeableness; C = Conscientiousness and Edu = Education. We use the convention that labels are printed at the positive side of the variable. Markers on the variable axes indicate standard deviation increases/decreases from the mean. Category points are labeled by the name of the drug together with a 1 (yes) or 0 (no) with D = Dysthymia, M = Major Depressive Disorder, G = Generalized Anxiety Disorder, S = Social Phobia (SP), and P = Panic disorder (PD).}
\label{fig:nesda1}
\end{center}
\end{sidewaysfigure}

The graphical display for the distress-fear model (Theory 2) is given in Figure \ref{fig:nesda1}. 
We can see that three response variables pertain to
the horizontal dimension, while two pertain to the vertical dimension. The class points for dysthymia, major depressive disorder, and generalized anxiety disorder fall on the horizontal dimension (and thus have vertical decision lines), while the class points for social phobia and panic disorder fall on the vertical dimension (and therefore have horizontal decision lines). The decision lines partition the 
two-dimensional space into rectangular regions in which a certain response profile is most probable. 
We further see that, on the horizontal dimension, dysthymia is best
discriminated as the two points lie farthest apart (distance 0.80), while
the distance for major depressive disorder is 0.64 and for generalized
anxiety disorder 0.56. On the vertical axis, social phobia is well
discriminated (distance 0.76) while panic disorder is hardly
discriminated (distance 0.16). The latter means that, using the three
background variables and the five personality variables together with the imposed model structure, we have hardly
any information to distinguish participants with and without panic
disorder. We will come back to this issue later.

The implied logistic regression coefficients are given in Table \ref{tab:nesda.coef}. As in our previous analysis we standardized the predictor variables such that the coefficients give changes in log odds for one standard deviation changes in the predictors. The most important predictor for mental disorders is neuroticism, which concurs with the conclusion in \cite{spinhoven2009role}

\begin{table}[t]
\centering
\begin{tabular}{lrrrrr}
  \hline
  & \multicolumn{5}{c}{Response Variable} \\ 
Predictor Variable & D & M & G & S & P \\ 
  \hline
Gender & 0.08 & 0.07 & 0.06 & -0.09 & -0.02 \\ 
  Age & 0.19 & 0.15 & 0.13 & -0.15 & -0.03 \\ 
  Education & -0.15 & -0.12 & -0.10 & -0.21 & -0.04 \\ 
  Neuroticism & 0.46 & 0.37 & 0.32 & 0.62 & 0.14 \\ 
  Extraversion & -0.27 & -0.22 & -0.19 & -0.24 & -0.05 \\ 
  Openness & -0.03 & -0.02 & -0.02 & -0.01 & -0.00 \\ 
  Agreeableness & -0.15 & -0.12 & -0.11 & 0.06 & 0.01 \\ 
  Conscientiousness & -0.09 & -0.07 & -0.07 & 0.14 & 0.03 \\ \hline
  Quality & 0.98 & 0.90 & 0.85 & 0.99 & 0.17 \\
   \hline
\end{tabular}
\caption{Implied logistic regression coefficients for the Depression and Anxiety data. D = Dysthymia, M = Major Depressive Disorder, G = Generalized Anxiety Disorder, S = Social Phobia (SP), and P = Panic Disorder (PD). The last row represents the quality of representation for the five response variables. }
\label{tab:nesda.coef}
\end{table}

The quality of representation for the five response variables is given in the last row of Table \ref{tab:nesda.coef}, where we see that the response variable panic disorder is poorly represented in this model. This could already be inferred from the graphical representation (the two points almost coincide) and the implied coefficients table, where most coefficients for the response variable panic disorder are very small. Apparently, when we use a standard logistic regression model for this response variable, the predictor variables discriminate the two categories much better. 

Because one response variable is poorly represented in the best fitting model, we also fitted an unconstrained model in two dimensions. Such a model has 28 parameters; the value of the loss function (deviance) is 4521.81 (AIC = 4578; BIC = 4708). The AIC indicates a better fit than the previous constrained models. The quality of representation of the response variables in this model is 0.97, 0.88, 0.85, 0.86, and 0.90, no longer indicating any poorly fitting response variables anymore. 

\begin{sidewaysfigure}
\begin{center}
\includegraphics[width = .95\textwidth]{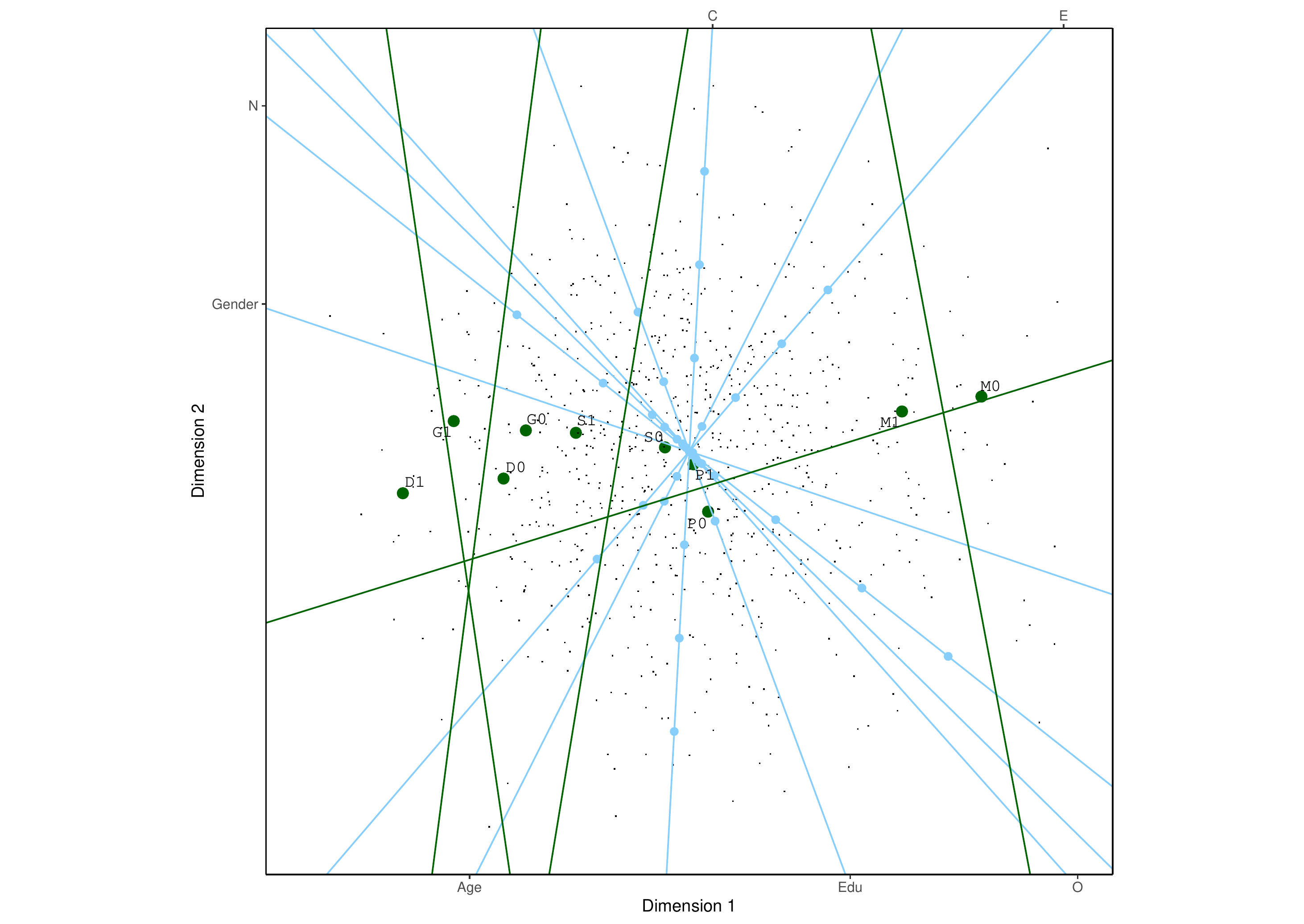}
\caption{Graphical representation of the two dimensional unconstrained solution for the depression and anxiety data. Predictor variable labels are printed on the border of the graph where N = Neuroticism; E = Extraversion; O = Openness to experience; A = Agreeableness; C = Conscientiousness and Edu = Education. We use the convention that labels are printed at the positive side of the variable. Markers on the variable axes indicate standard deviation increases/decreases from the mean. Category points are labeled by the name of the drug together with a 1 (yes) or 0 (no) with D = Dysthymia, M = Major Depressive Disorder, G = Generalized Anxiety Disorder, S = Social Phobia (SP), and P = Panic Disorder (PD).}
\label{fig:nesda2}
\end{center}
\end{sidewaysfigure}

The biplot for this two-dimensional unconstrained model is given in Figure \ref{fig:nesda2}, where we can see that the decision lines for major depressive disorder, dysthymia, generalized anxiety disorder, and social phobia run more or less parallel to the vertical dimension and  the decision line for panic disorder is more or less horizontal. This exploratory finding suggests a new theory: that major depressive disorder, dysthymia, generalized anxiety disorder, and social phobia pertain to a single underlying dimension, but that panic disorder behaves differently. 

\section{Related and competing approaches}

The MELODIC family is a statistical toolbox for simultaneous logistic regressions in a reduced dimensional space for the analysis of multivariate binary data. Other statistical models have been proposed for such data; in this section we show some relationships and comparisons. The related approaches can be divided in two types of models: marginal models and conditional models. 


\emph{Marginal models} are like standard regression models, dealing in some way with the dependency among responses. In generalized estimating equations 
\citep[GEE;][]{liang1986longitudinal, zeger1986longitudinal}, a working correlation structure is adopted and estimation and inference are adjusted based on this structure using a sandwich estimator \citep{white1980heteroskedasticity}.  GEE has been mainly developed in the longitudinal context but can also be applied to multivariate responses. Maximum likelihood estimation is also possible for marginal models -see for example \cite{bergsma2009marginal}- but this is computationally more demanding. 

Our MELODIC family can be seen as a member of the GEE family, where implicitly we adopt an independent working correlation structure. Moreover, we lay a dimensional structure on the response space. A standard GEE model would be equal to our one-dimensional model, with the constraint that all responses are equally well discriminated \citep[see][]{worku2018multivariate}. \cite{ziegler1998generalised} discuss a set-up where the predictor variables have different effects on each of the response variables. This set-up would correspond to our model in maximum dimensionality ($R$), where each response pertains to a single dimension. \cite{asar2014flexible} propose a method in which selected predictor variables have the same effect on selected response variables. This is similar to our constrained model, where predictors have a similar effect on response variables pertaining to a given dimension. 

As discussed in Section \ref{general-model}, our general MELODIC model is equivalent to reduced rank vector logistic regression models \citep{yee2003reduced}. Whereas we started our development from a distance perspective, these reduced rank models start from an inner-product perspective. \cite{derooij2005graphical} give an extensive discussion of the interpretational differences of the two perspectives. Reduced rank logistic regression models can also be visualized using biplots, as has been shown by \cite{vicente2020external}. The interpretation of these models is through projections on calibrated variable axes as in general biplots \citep[see][]{gower1996biplots, gower2011understanding}. We also proposed a constraint model, where \emph{a priori} knowledge about the dimensional structure of the response variables can be incorporated. As far as we know, such constrains have not been proposed in logistic reduced rank models before. 

In \emph{conditional models}, latent variables are included in order to model the dependency among the responses. The main examples are generalized linear mixed models and latent class models. The family of GLMMs includes item response models and factor analysis models \citep{skrondal2004generalized}. When these are expanded with predictor variables, explanatory item response models \citep{deboeck2004explanatory} and structural equation models are obtained. Explanatory item response models have mainly been  developed for unidimensional latent variables. Some progress has been made with multidimensional item response models, but the underlying structure should be known a priori (as in our constrained models). Explanatory multidimensional item response models still need further development, partly because estimation is often quite troublesome in such models due to the intractable integral in the likelihood function \citep{tuerlinckx2006statistical}. 

Latent class models \citep{lazarsfeld1968latent, mccutcheon1987latent} have been developed for multivariate binary data including predictors \citep{vermunt2010latent}. In latent class models, as in GLMMs, the dependency among the response variables is modeled using a latent variable, which in this case is categorical. No dimensional structure is imposed underlying the outcomes, only a choice of the number of categories of the categorical latent variable. These latent class models often require a large sample size in order to obtain stable and reliable results \citep{gudicha2016power}. 



\cite{hubbard2010gee}, when comparing generalized estimating equations and generalized linear mixed models, noted that 
``mixed models involve unverifiable assumptions on the data-generating distribution, which lead to potentially misleading estimates and biased inference''. More specifically, although the distribution of the random effects cannot be identified from the data, the estimates and inference change according to different choices of the distribution of these random effects. This makes the application of conditional models problematic. Another issue for conditional models is the number of indicator or response variables. In our example on depressive and anxiety disorders, for example, there are only five response variables. Distributing these five response variables over two underlying dimensions results in a dimension with only two dichotomous indicators. It is generally acknowledged that this number is much too low for valid inference. This small number is less of an issue in our family, because we do not assume a particular distribution for the underlying dimension. 

A final practical problem in these conditional models is that researchers often first try to find the dimensional structure and in a second step include the predictor variables. The measurement model (step 1), however, might change substantially when the predictor variables are included, leading to a completely different interpretation. To solve this problem, researchers have developed three-step \cite[][sometimes called the BCH approach]{bolck_croon_hagenaars_2004} and two-step estimators \citep{bakk2018two} within the context of latent class models which were recently adapted for other conditional models. In our family of models we have no division in measurement and structural model; the two go hand in hand.

\section{Conclusion and Discussion}

In this study, we presented distance models for simultaneous logistic regression analysis of multiple binary response variables based on ideas of multidimensional unfolding. Row objects (participants in our examples) are presented together with the two categories of the response variables in a low-dimensional Euclidean space, where the relative distance between a point representing a participant and the points representing the classes of a response variable determines the probability for each class. The model is estimated by minimizing a deviance function. 
These models take into account the dependency among the response variables by using a low-dimensional Euclidean space: with the increase in value of a predictor variable, the probabilities of all response variables change simultaneously. We christened the models the MELODIC family, that is the MultivariatE LOgistic DIstance to Categories family. We presented versions of the model both for cases in which  we have an \emph{a priori} theory about the dimensional structure of the response variables and for cases when we do not have such a theory. Two empirical applications are shown, one with and one without such an \emph{a priori} structure. 

In the case of a two-dimensional model, the result can be interpreted using a biplot. In the case of a higher-dimensional solution similar biplots can be constructed for pairs of dimensions. A coherent interpretation of the complete model from such bi-dimensional plots, might, however, more difficult. Alternatively, the model can be interpreted by the implied logistic regression coefficients which have a change in log odds interpretation similar to ordinary logistic regression models. We illustrated both methods of interpretation in the empirical examples. The fact that the model can be interpreted using a graph and using tables is beneficial, because applicants of statistical models can be divided into two groups: those who prefer visualizations and those who prefer numbers. With the MELODIC family, an applicant can choose which mode of interpretation is most suitable. 

We developed a fast iterative majorization algorithm to estimate the parameters of the model. The algorithm converges monotonically to the global optimum of the deviance function. The algorithm alternates between 1) updating an auxiliary vector $\mathbf{a}$, which is simply obtained by taking an average, and 2) updating the regression weights ($\mathbf{B}$) and item discriminations ($\mathbf{K}$), which can be obtained from a generalized singular value decomposition.  All model parameters can be obtained from these updates. 

A measure of quality of representation for each response variable was also proposed, ranging between 0 and 1. A higher value implies only a small loss of fit with regard to a univariate logistic regression with the specific response variable and the same set of predictor variables. This measure can be used as a diagnostic tool to assess whether response variables are well represented by the model.  In our second application, we saw that for one response variable the quality of representation was low. Further exploratory analysis suggested a different substantial theory.  

In applications, we need to select the predictor variables as well as the dimensionality of the model. We used information criteria such as the AIC and BIC in the empirical applications. Alternatively, cross validation or other model selection criteria can be used. We did not discuss uncertainty estimation for our model. Assuming the model is true, we can compute the Hessian matrix and derive standard errors for the parameter estimates from this matrix. Following the GEE set-up, we could develop a sandwich estimator for the covariance matrix of the parameters. Alternatively, the  bootstrap can be used \citep{efron1986bootstrap}. The two latter approaches acknowledge the fact that the model is an approximation \citep{buja2019amodels, buja2019bmodels} and probably not a completely accurate representation of a population model. In that sense, the sandwich estimator and the bootstrap can estimate uncertainty with respect to a target model in the population. Focusing on predictive accuracy instead of explanatory value \citep[cf.][]{Shmueli2010}, the performance of our model in comparison to that of independently fitted logistic regression models is expected to be higher \citep{breiman1997predicting}.  

In this manuscript we only focused on linear effects of the predictor variables. Non-linear effects of predictor variables on the responses can easily be incorporated as long as they can be translated into a design matrix $\mathbf{X}$, such as with the use of quadratic and cubic effects or with splines defined in terms of a truncated power basis \citep{friedman2001elements}. Non-linear variable axes can be presented in the graphical representation as smooth curves, where effects are still additive. Interactions can also be included in the model. In the graphical representation, \emph{conditional variable axes} need to be represented. In the case of an interaction between variables $X_1$ and $X_2$, the graphical representation has a variable axis for $X_1$ for \emph{each} value of $X_2$ (or the other way around). For an example of biplots with such interactions among predictors, see \cite{derooij2011transitional}. Note that both the nonlinearity and the interactions are effects with respect to all response variables. 

The past two decades have seen a rise in penalized estimation methods, which are methods that impose penalties on the parameters of the model. For the MELODIC family, these could be penalties on the regression weights to generalize the model to the case where $P >> n$, such as $L_1$ (Lasso penalty; \cite{tibshirani1996regression}) or $L_2$ penalties (Ridge penalty; \cite{hoerl1970ridge}) . To implement these in the MELODIC family, we would need to alter the identification constraints. In the current algorithm we used $n^{-1}\mathbf{B}^\top\mathbf{X}^\top\mathbf{X}\mathbf{B} = \mathbf{I}$ to identify the model, but this scaling does not seem to be in line with a penalty on $\mathbf{B}$. Therefore, the fixed scaling normalization should be placed on the discrimination values ($\mathbf{K}$).  Another potential type of penalty is a nuclear norm penalty \citep{fazel2002matrix}. In the outlined algorithm (see Algorithm \ref{melodic.alg}) we use a
singular value decomposition. If we apply an $L_1$ penalty to the singular values, the discrimination ($\mathbf{K}$) between the categories of response variables slowly diminishes. Because the singular values are ordered, the discrimination in the higher dimensions becomes zero first and later the discrimination of the first dimensions also becomes zero. If we choose the optimal value of the penalty parameter by cross validation, such a penalty could be used for dimension selection 

We are currently building an R-package that enables empirical researchers to apply the models to their own data. For the moment, the R-code of the examples presented can be found on github
(https://github.com/mjderooij/melodic).

\end{document}